\documentstyle[12pt,aasms4]{article}       

\input epsfig.sty      
\def\ltorder{\mathrel{\raise.3ex\hbox{$<$}\mkern-14mu 
             \lower0.6ex\hbox{$\sim$}}} 
\def\lsim{\lower.5ex\hbox{$\; \buildrel < \over \sim \;$}} 
                                                
\begin{document}      
       
% \thesaurus{% 06        % A&A Section 6: Form. struct. and evolut. of stars      
%             (08.14.1:  % Stars: neutron      
%              08.16.6)} % Pulsars: general      
       
\title{The converging inflow spectrum is       
an intrinsic signature for a black hole:      
Monte-Carlo simulations of Comptonization on free-falling electrons}      
       
 \author{Philippe  Laurent\altaffilmark{1} 
and Lev Titarchuk\altaffilmark{2,3}}      
       
\altaffiltext{1}{CEA/DSM/DAPNIA/SAp, CEA Saclay, 91191 Gif sur Yvette, France;  
fil@mir.saclay.cea.fr}   
\altaffiltext{2}{Laboratory for High--Energy Astrophysics,      
NASA Goddard Space Flight Center, Greenbelt, MD 20771, USA;      
titarchuk@lheavx.gsfc.nasa.gov}      
\altaffiltext{3}{George Mason University/Institute for      
Computational Sciences and Informatics, Fairfax VA}      
       
\date{Received ......., Accepted ........}         
       
\begin{abstract}      
An accreting black hole is, by definition, characterized by the drain.       
Namely, the matter falls  into a black hole much the same way as water      
disappears down a drain - matter goes in and nothing comes out.       
As this can only happen in a black hole, it provides a way to see 
``a black  hole", an unique observational signature of black holes.       
The accretion proceeds almost in a free-fall manner close  
to the black hole horizon, where the strong       
gravitational field dominates the pressure forces.      
In this paper we calculate (by using Monte-Carlo simulations)       
the specific features of X-ray spectra        
formed as a result of upscattering of the soft (disk) photons       
in the converging inflow (CI) within about 3 Schwarzschild radii of  
the black hole.   
The full relativistic treatment       
has been implemented       
to reproduce these spectra. We show that spectra in the soft state of       
black hole systems (BHS) can be described  as the sum      
of a thermal (disk) component and the convolution of some fraction of      
this component with the CI upscattering  spread (Green's) function.      
The latter boosted photon component is seen      
as an extended power-law at energies much      
higher than the characteristic energy of the soft photons.       
We demonstrate the stability of the power spectral index       
($\alpha= 1.8\pm 0.1$) over a wide range of the plasma temperature       
$0-10$ keV and mass accretion rates (higher than 2 in Eddington units).      
 We also demonstrate that the sharp high energy cutoff occurs       
at energies of 200-400 keV which are related to the average energy of      
electrons $m_ec^2$ impinging upon the event horizon.      
The spectrum is practically identical to the standard  
thermal Comptonization      
spectrum (Hua \& Titarchuk 1995) when the CI plasma temperature       
is getting of order of 50 keV (the typical ones for the hard state of BHS).      
In this case one can see the effect of the bulk motion only at 
high energies  where there is an excess in the CI spectrum with respect 
to the pure thermal one. 
Furthermore we demonstrate that the change of spectral shapes  
from the soft X--ray state to the       
hard X--ray  state is clearly to be related       
with the temperature of the bulk flow. In other words        
the effect of the bulk Comptonization compared to the thermal one  
is getting  stronger when the plasma temperature drops below 10 keV.       
These Monte-Carlo simulated CI spectra are a inevitable stamp of       
the BHS where the strong gravitational field dominates        
the pressure forces.

\end{abstract}      
       
\section {Introduction}      
      
A lot of theories and models have been suggested to       
explain the observational situations of black hole systems (BHS);      
for recent reviews of the astrophysics of black holes [see e.g.,      
\cite{liang97}, \cite{zhang97}, \cite{tit96}, \cite{tit97}].      
      
The BHS are generally observed       
in two spectral states: one  with a high luminosity soft thermal bump       
and a low luminosity power law extending to several hundred of keV       
(hereafter called the ``soft state"), and the other at high luminosity       
in  high energies ($> 10$ keV) showing a typical thermal Comptonization       
component (called the ``hard state" all along the paper).      
      
It is now clear  that the  observed X-ray spectra  resulting
from the accretion of
matter onto the central object is divided
according to the connection between the two effects:
(i). The gravitational energy of matter  is released in the disk due
to the viscous dissipation between the adjacent  layers
(Shakura \& Sunyaev 1973,
hereafter SS73). The energy release occurs
in the optically thick medium
and thus the disk emits a soft blackbody like radiation
(related to the effective disk area, SS73) of a fraction of
keV and tens of eV for galactic and extragalatic sources respectively.
(ii). Another mechanism is related with the geometric compression of
the matter accreting in advection dominated manner
where the radial velocity is some fraction of the free fall velocity
but the viscous heating can be  essential in the flow too  
(e.g. Chakrabarti \& Titarchuk 1995, hereafter CT95; Narayan \& Yi 1994). 
Energy is mostly transferred by the protons due to their much higher mass.
These protons afterward transfer their energy to the ambient electrons through  
Coulomb collisions in a relatively dense cloud (called the Compton cloud)  
close to the black hole.  
This cloud can be a result of the propagation (accretion) of  
the sub-Keplerian component towards the black hole. In such a 
type of accretion, shocks or centrifugal barrier regions can be formed (CT95). 
      
Since the cross-sections for radiative processes are inversely       
proportional to the mass of the emitter, it is the electrons       
which then radiatively cool the plasma. In such an high temperature plasma, 
the main electron cooling process is the Comptonization on the ambient       
soft photons. The plasma temperature of the zone emitting 
the hard radiation is  regulated by the supply of soft photons from the disk.
       
It was demonstrated in  CT95  that the hard-soft state transition in BHS 
is regulated by redistribution of mass accretion rates between the 
quasi-spherical (sub-Keplerian) and the Keplerian disk components.      
It was also shown there for the first time that 
the converging inflow  spectral feature 
(as an intrinsic signature of black hole) can be seen in      
the soft state. Namely, near the horizon, the strong gravitational field 
is expected to dominate the pressure forces and thus 
to drive the accreting material in almost a free fall i.e. a converging 
inflow (Titarchuk \& Zannias 1998, hereafter TZ98). 
Unlike, for other compact objects the pressure forces
are becoming dominant as their surface is approached, and thus such a free 
falling state is absent. 
In contrast CT95 emphasized that in the soft state, the optically      
 thick Compton cloud surrounding a black hole is cooled down by the disk  
soft photons and the extended power law will be observed.         
 
The fact of the presence of the geometrically thin  
disk in BH system leads to the very important statement regarding the 
dominance of gravitational forces over radiative forces:  
{\it Because the gravitational force in the disk is 
a small $H/R$ fraction of the gravitational force acting 
in the radial direction 
(where $H$ and $R$ are the disk height and radius respectively) 
the local luminosity in the disk cannot be higher than $H/R$ 
times the Eddington limit}. 
 At some certain point the disk is disrupted or terminated, 
for example due to General Relativistic effects: 
the  stable orbits do not exist below  three  
Schwarzschild radii. Then the matter is forced to proceed towards the black
hole in almost free-fall manner - the radiative forces 
being negligible with respect to the gravitational forces. 
The next question now is to know how the soft disk photons illuminating 
the bulk inflow can be upcomptonized to energies of the order of $m_ec^2$.  
 
Titarchuk \& Zannias 1998 show that the production of hard photons as a 
function of the radius from the black hole peaks at around 2 Schwarzschild 
radii ($2r_s$) and most of the hard continuum photons observed are 
produced within 3$r_s$ from the black hole.   
Shrader \& Titarchuk 1998 (hereafter ShT98) 
demonstrate that the emergent spectra 
of the converging inflow atmosphere successfully explain the 
continuum X-ray spectra of black hole systems (BHS) in their soft state.   
      
To check these statements and predictions we have simulated the emergent
energy spectrum, treating as precisely as possible the Comptonization process.  
To do this, we need to know first what is the movement of the high 
energy electrons: it is in fact, a composition of the free-fall motion onto the black hole and of the Brownian (thermal) motion which can be characterized  
by a temperature.       
Whereas the thermal Comptonization has been already explored 
in several ways (e.g. Sunyaev \& Titarchuk 1980; 
Pozdnyakov,  Sobol' \& Sunyaev 1983, 
hereafter PSS; Nagirner \& Poutanen 1994; Titarchuk 1994, hereafter T94;  
Titarchuk \& Lyubarskij 1995; 
Hua \& Titarchuk 1995; Giesler \& Kirk 1997), 
the effects of the free-fall motion of       
the electrons on the resulting spectrum is not yet clearly understood.       
Also, the general relativistic effects close to the black hole were not 
yet plainly taken into account.       
In series of papers, Titarchuk, Mastichiadis \& Kylafis (1996), (1997)      
(hereafter TMK96 and TMK97) present the  exact numerical and approximate      
analytical solutions of the problem of spectral formation 
in a converging inflow, taking into account the inner boundary condition, 
the dynamical effects of the free fall, and the thermal motion of 
the electrons.      
The inner boundary has been taken at a {\it finite} radius with 
the spherical surface being considered as a fully absorptive.      
      
TMK  have  used a variant of the  Fokker-Plank formalism      
where the inner boundary mimics a black-hole horizon;      
no relativistic effects (special or general) have been taken into account.      
Thus their  results are instructively useful but  they are not directly 
comparable with the observations.      
{\it But using these numerical and analytical techniques  
they demonstrated that the extended power laws are present 
in the resulting spectra in addition to the blackbody like emission 
at lower energies.}       
      
Our approach is to determine a realistic emergent Comptonization spectrum       
 using two different approaches: one, purely analytical, is described in      
detail in  TZ98, and the other, numerical, is presented below.      
TZ98 analyzed the exact general relativistic integro-differential      
equation of radiative transfer describing the interaction of low energy 
photons with a Maxwellian distribution of the relativistic electrons 
in the gravitational field of a Schwarzschild black hole. 
They  proved that due to Comptonization an initial arbitrary  
spectrum of low energy photons unavoidably results in spectra characterized 
by an extended power-law feature.      
They examined the spectral index $\alpha$ by using both analytical  
and numerical methods  in the lower temperature limit 
($\Theta=kT_e/ m_ec^2\ll1$), where $T_e$ is the electron temperature and 
$m_ec^2$ the electron rest mass energy, and with a varying mass accretion 
rate. They demonstrated the stability of the ($\alpha\sim1.8$) power-law,  
which is mainly due to the asymptotic independence of the spectral index  
on the mass accretion rate and its weak dependence on plasma temperatures 
in the lower temperature limit. In fact, all corrections to the index       
due the temperature effects are related with the corrections  
of the kernel of the relativistic kinetic equation (TZ98) which are  
of the order of the dimensionless temperature $\Theta$.        
      
In this paper, we will  describe the numerical approach        
briefly outlining the main features of the Monte-Carlo simulation code       
in \S 2.      
We will present the results in the ``Flat" case in \S 3       
for which  we assume the propagation of the photons in straight lines.       
Then, we will show what we have obtained in the most realistic       
general relativistic case in \S 4. In sections 3-4,        
we will compare also these results to the ones       
obtained by  different (analytical and numerical) methods in the same   
physical framework (TMK97 and TZ98). 
We will show that the observed spectral transitions in BHS is clearly 
to be related to the bulk flow temperature in
 \S 5.  We will demonstrate the relevance of the converging inflow model
 to the recent high energy observations in \S 6.
  Finally, we will summaries our work  and draw conclusions in \S 7.       
%Figure 1 ---------------------------------------      
% 
%\begin{figure*}      
%\centerline{\epsfig{figure=fig1f.ps,height=200mm}}      
%\caption{The left-hand panel is an artists 
%conception of an accretion disk surrounding a black hole. The 
%approximate sites from which the soft-seed radiation, the more general 
%soft radiation which is directly viewed by the observer, and the hard- 
%radiation emanate are indicated. Also, the dotted lines indicate the 
%scale and location of the bulk-motion scattering site. The schematic  
%diagram 
%in right-hand panel, depicts a typical sequence of scatterings within the 
%bulk-motion inflow region. Some details are described in the text.}      
%\label{Figure1}      
%\end{figure*}      
 
\section{The Monte-Carlo simulation}      
As explained above, this signature originates from upscattering of low 
energy photons by fast moving electrons  with velocities, $v$, 
approaching the speed of light, $c$. A soft photon of energy $E$, 
 in the process of multiple scattering off  the electrons, gets 
substantially blue-shifted. For a single scattering,
the photon scattered off an electron moving with velocity $v$, 
changes its energy due to Doppler effect as follows 
\begin{equation} 
E^{\prime}=E{{1-(v/c)\cos\theta}\over{1-(v/c)\cos\theta^{\prime}}}. 
\end{equation} 
\noindent 
The photon is then significantly blue-shifted if, at least, it gets once
scattered in the direction of the electron motion 
 (i.e. when $\cos\theta^{\prime}\approx 1$).  
This is schematically illustrated in Figure 1 for two scattering events. 
In the first event the soft photon is effectively upcomptonized and in the  
second one the photon is scattered in such a direction that it can be 
detected by an observer. 
 
{\bf Editor, please put Figure 1 here.} 
  
In the first scattering event we assume the direction of 
incident photon,  $\theta_1$, to be nearly normal to the electron velocity, 
 and the direction of the scattered photon to be nearly aligned with the 
electron velocity. During its outward propagation through the 
converging-inflow medium, the angle between the photon and electron 
velocity increases. Thus, in the second event the cosine angle, $\cos 
\theta_2$, 
tends to approach zero. The angle of outgoing photon, $\theta_2^\prime$, 
has to be large enough,  in order for the Doppler boosted photon 
to reach an observer.               

The geometry mostly used in these simulations is shown in Figure 2:       
it consists of a thin disk surrounding a spherical cloud harbouring       
a black hole in its center. The disk is always supposed to be optically       
thick,  whereas we are assuming a free fall for the background flow where       
the bulk velocity of the infalling plasma is given  
by $v(r)=c(r_s/r)^{1/2}$.      
 
{\bf Editor, please put Figure 2 here.} 

%\begin{figure*}      
%\centerline{\epsfig{figure=fig2f.ps,height=50mm}}      
%\caption{Scheme of the geometrical setup used       
%in the Monte-Carlo simulations. We vary the accretion sphere radius  
%$r_{out}$ from  3 $r_s$ to 20 $r_s$ in our simulations (see text).}      
%\label{Figure2}      
%\end{figure*}      
In our simulations (see also the kinetic formalism in TZ98) we use      
 the number density  $n$ measured in      
the local rest frame of the flow that is       
$n=\dot m(r_s/r)^{1/2}/(2r\sigma_T )$. Here $\dot m=\dot M/\dot M_E$,      
$\dot M$ is the  mass accretion rate, $\sigma_T $ is the Thomson cross section,      
  $\dot M_E \equiv L_E/c^2=4\pi GMm_p/ \sigma_Tc~$ is      
the Eddington accretion rate and $M$ is the black hole mass.      
Then the optical thickness of the cloud      
is computed through  the mass accretion rate $\dot m$ according to the       
following formulae :      
\begin{equation} 
\tau = \dot m \left( 1 - \sqrt{r_s \over r_{out}} \right )       
\end{equation} 
\noindent 
where $r_{out}$ and $r_s$ are the outer boundary and the       
Schwarzschild radii, respectively.  
In most of all results given below, except otherwise presented in part 4, 
 the cloud radius $r_{out}$ has been taken to be 3 $r_s$.  
The different values of $\dot m$  which has been used  
in those simulations are       
$\dot m = 0.5,1,1.5,2,4,7$.      
      
We have varied  the cloud electron temperature from 0 to 50 keV.       
In addition to the thermal motion of the electrons in the cloud,       
we have also taken into account their free-fall onto the central black hole. 
The seed X-ray photons were generated uniformly and isotropically       
at the surface of the inner edge of the accretion border of the disk.       
For our simulations we  use as an example the disk thermal spectrum       
with a temperature of 0.5 keV.        
      
In the simulations, we have followed the trajectory of each photons       
in the following way: first, we have made a random generation       
according to an exponential law: $P(\tau) = e^{-\tau} \;$ 
to get the value of the optical thickness $T$ that a photon will 
travel before interacting. 
Then we have integrated the optical thickness  
$\tau = \rho \sigma l$ along the photon path up to $T$,  
taking into account the variation of the cloud density $\rho$ and  
of the cross section $\sigma$ with the radius.  
The gravitational shift endured by the photon was also computed at each  
step of this integration.      
      
The photon paths are straight lines in what we called the ``Flat" case,       
or curves in terms of  the Schwarzschild geometry in the general relativistic 
case.       
In fact, the variation of the photon energy, and of the photon position 
(with respect to the local normal) along the photon trajectory can be extracted 
from the full kinetic equation presented in TZ98 (Eq. 12).       
The characteristics of the differential part of the equation provides       
these dependencies.      
 
If, at the end of the integration, the photon has not left the cloud,  
we simulate a Compton scattering with an electron according       
to the method described in PSS, taking into account the exact motion  
of the electron, that is a composition of its free-fall motion onto  
the black hole with its Brownian thermal motion.  
To do this, we first compute the scattering electron momentum,  
and derive the scattered photon and electron characteristics  
from the Compton scattering kinetics.  
We then check if this event is consistent with the Compton  
scattering probabilities;  
if yes the event is kept; if not another scattering electron is generated,  
and the process goes on until the event is accepted.  
This process has been successfully checked by comparing its results  
with the analytical ones of Hua \& Titarchuk 1995 in the case of relativistic  
thermal electrons. 
     
Once the new energy and direction of the photon been determined by the  
Compton kinematics, we track it in the same way as above until it makes  
another scattering, or it escapes from the cloud, or until it is ``absorbed"  
by the black hole at its horizon.       
      
The parameters of the simulations are the Compton cloud electron temperature,  
$T_e$, the cloud outer radius, $r_{out}$ and 
the mass accretion rate, $\dot m$.  
         
\section{ ``Flat" Case}      
      
We have first investigated what could be the emergent spectrum      
in the ``Flat" case, that is supposing that  photons goes in       
straight lines.  But the relativistic effects such as gravitational redshift, 
dependence of the Klein-Nishina cross section on energy,       
or ``absorption" of photons at the black hole horizon were correctly       
taken into account in these computations.  
Even if this case is not physically consistent,  
as we cannot separate in theory the gravitational  
redshift effects from the light bending, it is instructive as  
it will enable us to derive the consequences of space 
 curvature around the black hole on the emerging spectrum,  
by comparing this case to the general relativistic one described below.      
  
We have calculated the spectrum emerging from a plasma cloud of temperature 5 keV 
for $\dot m$ equal to 2 with an outer radius, $r_{out}=3 r_s$.      
They are typically the physical conditions of the soft state.       
As foreseen theoretically, (TMK96-97)       
the spectrum observed at infinity consist of a soft component       
(blackbody like) coming from input photons that escaped after a few       
scattering without any significant energy change and       
the high energy part (a power law)  extending to high energies of order       
300 keV. The shape of spectrum  is reminiscent of what is observed 
from BHS in their soft state (ShT98) 
 and it is almost identical to the spectrum presented       
in TMK97 (Fig. 1).      
      
However it is clear that even in the low temperature case,       
the spectrum we obtained  is too hard to account for what is observed  
from soft state BHS.      
  We have indeed fitted the Monte-Carlo       
spectra (see below Fig. 5)  
with a test spectrum, composed of two components:       
a Planckian spectrum with a fixed temperature of the soft photons and       
convolution of some fraction of      
this component with the converging inflow  upscattering  Green's function.      
The fit parameters were then the color temperature of the soft  
component, the normalization of the upscattering 
 component relative to the soft one,        
and the spectral index of the power law.  
An example of this fit to the spectrum shown in Figure 5 
(which is a spectrum obtained in the General Relativistic case, 
with $kT_e =$ 20 keV and $\dot m = 4$), is given in 
Figure 3. 
 
  The fitting parameters are $kT_c=0.4$ keV, the fraction  
of the illumination of the bulk inflow atmosphere by the disk photons,  
$f= 0.48$, and the spectral index $\alpha=1.5$.  
It is worth noting that the fitting temperature $kT_c=0.4$ keV 
measured by an observer is less than the input temperature of the soft  
photons 0.5 keV because of the gravitational redshift.  
 
{\bf Editor, please put figure 3 here. } 

%\begin{figure*}      
%\centerline{\epsfig{figure=fig3f.ps,height=100mm}}      
%\caption{The Monte Carlo spectrum (see below Fig. 5) 
% is fitted to the test spectrum, composed of two components:       
%a Planckian spectrum with a fixed temperature of the soft photons and       
%convolution of some fraction of      
%this component with the CI upscattering  Green's function}.      
%\label{Figure3}      
%\end{figure*}      
      
{\it So, as it could be seen in Table 1, 
the  spectral index we found by fitting the computed spectrum in 
the ``Flat" case, are around or harder than      
1, which is harder than  it has been observed in the soft state (around      
 1.5, see for instance Ebisawa, Titarchuk \& Chakrabarti, 1996,       
hereafter ETC). } It is worth noting that the spectral indices presented 
in table 1, are very close to TMK97 results (see Figs. 6 there) made       
with an assumption of a purely Newtonian geometry.       
      
So, as expected, the modelling we made in the ``Flat" case is       
not realistic.       
It has been shown moreover that it is possible       
to soften the emergent spectrum by taking into account       
general relativistic curvature effects (TZ98).       
 We have verified this result by computing the Comptonized spectrum       
in a fully general relativistic case, taking into account       
the space curvature around the black hole horizon.      
%Table 1 -------------------------------------------------------------------      
       
\begin{table}       
\caption{Spectral index of the power law tail, obtained in the ``Flat" case.}
      
\label{Table1}      
\[      
    \begin{array}{ccccccc}      
    \hline      
\noalign{\smallskip}      
      
\rm kT_e & \dot m = 0.5 & \dot m = 1 & \dot m = 1.5 & \dot m = 2 &       
\dot m = 4 & \dot m = 7 \\      
\hline      
\noalign{\smallskip}      
\rm 0 ~keV &  2.4^*   & 1.6^*  &  0.9   &   0.8   &   0.9   &  1.2 \\      
\hline      
\noalign{\smallskip}      
\rm 5~keV &   2.1^*   & 1.1  &  0.8   &  0.8    &   0.9   &  1.0 \\      
\hline      
\noalign{\smallskip}      
\rm 10 ~keV &  2.0^*  &  1.1  &  0.8   &  0.8  &  0.9  & 1.0 \\  
\hline      
\noalign{\smallskip}      
\rm 50 ~keV &  1.6  &  1.1   &   0.8   &  0.9   &  0.7    & 0.5  \\      
\hline    
   \end{array}      
   \]      
* The spectral index determination is less accurate for these values       
of parameters because the hard component of the spectrum is 
less pronounced in the fit region (10-50 keV).        
\end{table}         
       
\section{The General Relativistic Case}      
      
For this case we have  modified our simulation code in order to take into       
account the space curvature around the black hole horizon:       
instead of following the photon path along a straight line,       
we have computed at each step the exact curved trajectory of the photon.      
The other physical effects (Compton diffusion, redshift, electrons thermal
and free fall motion, {\it etc, ...}), has been computed in the same way as 
described previously. This description is then now fully consistent with  
the General Relativity theory. We can see in Figure 4 an example of       
the resulting emerging spectrum for $kT_e =$ 5 keV, and $\dot m = 2$:
as expected, this spectrum is softer than the one we obtained in the
``Flat" case (compare  table 1 and 2 and also see TMK97).  

{\bf Editor, please put fig. 4 here}

A summary of the results we calculated for different physical       
conditions is given in table 2, where we have put the spectral index       
of the power law, while we fit our spectrum with a two-component       
test spectrum as described above.        
The relative error on the quoted value of the spectral index in       
table 2 is always around $\pm 5 \%$. We can notice from this table       
that, as discussed in TZ98, the spectral index we found  
in the low $kT_e$ regime, are       
only slightly dependent of the mass accretion rate $\dot m$,       
provided that $\dot m$ is larger than 1.       
      
{\it Finally, we can see in table 2 that the index of       
the computed spectra in the soft state are now       
between 1.5 and 2.0, which is closest to the observations.}       

%Figure 4 ---------------------------------------      
%\begin{figure}      
%\centerline{\epsfig{figure=fig4f.ps,height=70mm}}      
%\caption{Emergent Comptonized spectrum in the general relativistic case:       
%in this simulation, the cloud has a temperature of 5 keV       
%and $\dot m$ is equal to 2.       
%The best power law fit also shown has a  spectral index of 1.9.}      
%\label{Figure4}      
%\end{figure}      
      
It is worth noting here that  the observed power law index of       
Narrow Line Seyfert 1 (NLS1) galaxy population -      
which may represent the extragalactic analogue of the BHS in  
the high-soft state - is around $1.8$.  
This  is close to what we have obtained       
in the low $kT_e$ regime,       
as it is expected from Comptonization on electrons whose motion is      
dominated by the bulk free-fall.       
       
%Table 2 ------------------------------------------------------------------- 

\begin{table}       
\caption{Spectral index of the power law tail, obtained       
in the general relativistic case.}      
\label{Table2}      
\[      
    \begin{array}{ccccccc}      
    \hline      
\noalign{\smallskip}      
      
\rm kT_e & \dot m = 0.5& \dot m = 1& \dot m = 1.5& \dot m = 2 &       
\dot m = 4 & \dot m = 7 \\      
\hline      
\noalign{\smallskip}      
\rm 0 ~keV & 2.8  & 2.3  &  2.1    & 2.05  &  2.0    &  1.9 \\      
\hline      
\noalign{\smallskip}      
\rm 5~keV & 2.8   & 2.2  &  2.0    &  1.9  &   1.8   &  1.8 \\      
\hline      
\noalign{\smallskip}      
\rm 10 ~keV &  2.8    &  2.1  &   1.9   & 1.9  &  1.7  & 1.7 \\      
\hline      
\noalign{\smallskip}      
\rm 50 ~keV &  1.8   &  1.4 &  1.2   &  1.0 &   0.9 &  0.7 \\      
\hline      
   \end{array}      
   \]      
\end{table}       
      
As  is seen from Figs 4-5, the power-law part of the spectrum occurs       
at energies $E$, lower than that of the exponential cutoff,  
$E_e$ (where $E_e$ is the average electron energy).       

{\bf Editor, please put fig. 5 here}

 TZ98  have exactly solved       
the radiative transfer equations in the Doppler regime
(when the photon energy is less then the energy of the electrons) 
in the same fully relativistic framework than the one we present here.      
Namely, they demonstrated that the Green's function is a broken  power law      
and they calculated exactly the high energy power index       
 in the low temperature limit, for $\Theta\ll 1$.      
      
We have compared our results and results of TZ98   obtained  
for zero temperature       
for 3 different geometry of the Monte-Carlo simulation.   
The geometry used are the followings:      
      
   1) a $3 r_s$ sphere with a monochromatic soft X-ray source       
($\rm E_p = 0.5$ keV) placed at the edge of the accretion disk,       
as shown in Figure 2.      
      
   2) The same as above but with a sphere radius of $20 r_s$.      
      
   3) a $20 r_s$  Comptonization cloud, with the source of 0.5 keV       
   X-ray photons placed inside in a ring at $5 r_s$.      
           
Firstly we found only small differences in the spectral indices       
for all three cases, i.e.      
differences which are all within the error range of the spectral index       
determination, being around 0.1 for $\dot m $ greater than 2-3.      
Secondly, we found the mean spectral index value is exactly       
$1.8$ in agreement with TZ98. We  also note  from this  
comparison that, as it is theoretically expected, the power law part  
of spectra is almost independent of the soft photons distribution. In  
other words, it is independent of the illumination of the bulk motion  
site by the accretion disk flow.          

%Figure 5 ---------------------------------------      
%\begin{figure}      
%\centerline{\epsfig{figure=fig5f.ps,height=70mm}}      
%\caption{Emergent Comptonized spectrum in the general relativistic case:       
%the physical conditions of the cloud are now: $kT_e =$ 20 keV, $\dot m = 4$.     
%The best power law fit also shown has a spectral index of 1.5.}      
%\label{Figure5}      
%\end{figure}      
     
In Figure 5 we give the result of our       
computations for $kT_e =$ 20 keV, and $\dot m = 4$       
which reproduces the main features of the spectra observed       
in the high state of BHS: a spectral index of $1.5$ and an high energy       
cutoff around 500 keV.       
The effect of the thermal motion for temperatures      
higher than 10 keV is there   
clearly seen. The spectrum is getting harder, i.e.        
the index decreases and the high energy cutoff increases.       
      
Furthermore, the spectra obtained at high $kT_e$ and $\dot m$ greater       
than 2 have a spectral index       
around $0.8$ which is what is observed from BHS in the hard state,       
where the thermal Comptonization is thought to dominate (ETC).       
      
This is also shown in Figure 6 where we present the results of the       
computation for $kT_e =$ 50 keV, and $\dot m = 4$,       
compared to  the analytical solution of Hua \& Titarchuk 1995 (Eq. 6)  
derived for the pure thermal Comptonization case.       
The spectrum is practically identical to the standard thermal Comptonization 
spectrum.      
In this case the effect of the bulk motion can be seen only  
at high energies  where there is an excess in the converging inflow spectrum  
due to coupling of the thermal and bulk motion  
velocities. It is worth noting here that this excess is really detected  
in the observation of the hard state of Cyg X-1 source  
(e.g. Ling et al. 1997, see also for details of the model,   
Skibo \& Dermer 1995).       
 
{\bf Editor, please put Figure 6 here.}       
      
So, the change of spectral shapes from the soft X--ray state to the       
hard X--ray  state is clearly to be related       
to the temperature of the bulk flow. In other words        
the effect of the bulk Comptonization compared to the thermal one  
is getting      
stronger when the plasma temperature drops below 10 keV.       
Also, as it could be seen by comparing the results in the ``Flat" and       
general relativistic case, 
the spectral index of the Comptonized spectrum is     
very dependent of space curvature effects, and 
then its observation should       
be a very powerful tool  to determine the nature of the central object.       
 
%Figure 6 ---------------------------------------      
%\begin{figure}      
%\centerline{\epsfig{figure=fig6f.ps,height=70mm}}      
%\caption{Emergent Comptonized spectrum in the general relativistic case       
%with this time $kT_e =$ 50 keV, and $\dot m = 2$.       
%The high energy part of the spectrum is compared       
%to the thermal Comptonization spectrum [Hua \& Titarchuk (1995), Eq. 6].}       
%\label{Figure6}
%\end{figure}      
 
\section{Spectral State Transition and  Temperature of the Converging  
Inflow Atmosphere} 
 
To complete the discussion of the Monte Carlo  
simulations and their application  to the observed spectra of BHC systems we  
present some details of the thermal balance treatment of  
the bulk motion atmosphere (cf. Chakrabarti \& Titarchuk 1995, 
Titarchuk, Lapidus \& Muslimov 1998). 
  
The total count rate from the source increases with an increasing  
accretion rate.  
The X-ray spectrum becomes softer, due to the increase of the supply of  
soft photons from the disk illuminating the advection dominated  
region (shocks, centrifugal barrier) where the energy release is  
due to geometrical compression of the non-Keplerian accreting matter. 
This region, the so called Compton cloud, is the source of  
hard X-ray radiations which are produced as a result of the upscattering of  
the soft photons emitted from the disk. In the final stage of the accretion 
(this is within a few Schwarzschild radii)   
the matter goes towards the black hole almost free falling as 
the radiation pressure cannot stop matter falling in. 
 
The Compton cloud (CC) region can be treated as a potential wall at which the 
accreting matter releases its gravitational energy. {\it This occurs 
in an optically thin region where the column density is of order of a 
few grams}, or where the Thomson optical thickness $\tau_0 \sim $ a 
few. The amount of energy released per second is a fraction of the Eddington 
luminosity since the CC region is located in the very vicinity of 
a central object ($\sim $ 3-6 $\rm r_s$). The heating of a gas due to 
the gravitational energy release should be balanced by the photon 
emission. For high gas temperatures, Comptonization is the 
main cooling channel, and the heating of electrons is due to their   
Coulomb collisions with protons. Under such physical conditions the 
energy balance can be written as (see e.g. Zel'dovich \& Shakura 1969, 
hereafter ZS69, equation [1.3]) 
\begin{equation}      
 F/\tau_0 ~=~ C_0\cdot \varphi(\alpha) 
\varepsilon(\tau)T_e/f(T_e) . 
\end{equation} 
\noindent 
Here $\tau$ is the current Thomson optical depth in the emission 
region (e.g. in a slab), $\alpha $ is the energy spectral index for   
a power-law component of the Comptonization spectrum,  
$\varepsilon(\tau)$ is a distribution function for the radiative  
energy density, $\rm f(T_e) = 1+2.5(kT_e/m_ec^2)$, $\rm T_e$ is the  
plasma temperature in K, $\rm C_0 = 20.2$ cm s$^{-1}$ K$^{-1}$ is a  
dimensional constant, and, finally, $\rm \varphi (\alpha ) = 
0.75\alpha (1+\alpha /3)$ if $\varphi (\alpha )\leq 1$, otherwise  
$\varphi (\alpha ) = 1$. The latter formula is obtained by using  
the relationship between the zero- and first-order moments (with  
respect to energy) of the Comptonized radiation field  
(Sunyaev \& Titarchuk 1985, \S 7.3, equation [30]). The distribution  
of the radiative energy density in the emission region  
$\varepsilon (\tau )$ can be obtained from the solution of  
the diffusion equation (cf. ZS69, equation [1.4]),  
\begin{equation}      
 {1\over3}{{d^2 \varepsilon}\over{d\tau^2}}=-{(F/c)\over{\tau_0}} , 
\end{equation}      
\noindent 
subject to the two appropriate boundary conditions.  
 
\noindent 
The first boundary condition must imply that there is no 
scattered radiation from the outer side of the emission region, i.e. 
\begin{equation}      
\rm {{d\varepsilon}\over{d\tau}}-{3\over2}\varepsilon=0~~~for~\tau=0.  
\end{equation}      
\noindent 
In our case this condition holds at the inner surface of a slab facing 
a central object (black hole). 
In the neutron star case one can expect some 
additional soft flux from a NS resulting in an additional illumination  
of the inner surface of a slab. It is easy to generalize  
 the following analysis for this particular case and we offer the reader  
to make this exercise.  
 
\noindent 
The second boundary condition requires that at the outer surface of a 
slab the incoming flux should be equal to the external flux $F_d$ 
 i.e.  
\begin{equation}      
{1\over3}{{d\varepsilon}\over{d\tau}}={F_d\over c} ~~~at~\tau=\tau_0. 
\end{equation}      
\noindent 
 
The solution of equations (4)-(6) provides us with the distribution  
function for the energy density  
\begin{equation}      
\rm \varepsilon(\tau)= {{F+F_d}\over{c}} 
\{2+3\tau_0[\tau/\tau_0-0.5(\tau/\tau_0)^2 F/(F+F_d)]\}. 
\end{equation}      
\noindent 
Thus, from equations (3) and (7) we get 
\begin{equation}      
{{\varphi(\alpha)T_e\tau_0}\over{f(T_e)}}\ltorder  
0.75\cdot 10^9{{F}\over{F+F_d}}~{\rm K}.  
\end{equation}      
\noindent 
When $\rm F_d\ll F$ the spectral index $\alpha $ varies very little since 
$\alpha $ is a function of $\rm T_e\tau_0/f(T_e)$ (see e.g.  
Titarchuk \& Lyubarskij 1995 for the thermal Comptonization case). 
Thus,  
as long as the external 
flux (due to the photons from the disk) is much smaller than the 
internal energy release (per cm$^2$ per second) in the CC region, the 
spectral index is insensitive to the accretion rate in the disk.  The 
values of parameters consistent with equation (8) are: 
$\rm \tau_0\lsim 5$, $\rm T_e\lsim 2\times 10^8$ K, and 
$\alpha \lsim 1$, which are characteristic of a hard state for the  
galactic BH and NS systems. 
  
When the external flux $F_d$ becomes comparable to the internal energy 
release, F, the cooling becomes more efficient not only due to 
Comptonization, but also due to the free-free cooling, and therefore 
the electron temperature unavoidably decreases, $\rm T_e\ll 10^9\cdot 
F/(F+F_d)$ K (see CT95 for the numerical calculations of spectral 
indices and temperature).  
 
This illumination effect can  explain the  
hard-soft transition when the temperature of the Compton cloud drops 
substantially with the increase of the soft photon flux  
from the disk. As it is seen from above (Eq. 8)  
the temperature drops from 50-60 keV in the hard state 
(i.e. when $F_d\ll F$) to 5-10 keV in soft state (when $F_d$ and $F$  
are comparable). For this particular range of temperatures 
the Monte Carlo simulations have been done in this paper.  
       
\section{The relevance of Bulk motion Comptonization model to the recent
high energy observations}      

Shrader and Titarchuk, (1998), have already shown that the spectral shape 
of Compton
scattered photons in converging flow is very close to the observed spectra.
However, it is still  necessary to show how many photons {\it escape} 
from the central region, when a given number of photons are injected 
in the Monte-Carlo simulations.  
For example in the case of $kT_e = 20$ keV, and $\dot m = 4$ (see Fig.5), 
among  $10^4$ generated photons there are 
15,372 photons, or  15.4 \%, escaping without interacting,
45,557 photons, or  45.6 \%, going through the black hole horizon,
13,393 photons, or 13.4 \%, scattered in the accretion disk (but not in 
the converging Compton cloud), the inner radius temperature of which 
$=0.5$ keV, 
and 25,678 photons, or 25.7 \%, scattered in the Compton cloud (some of 
these photons being also scattered in the disk).
Thus almost the same photon flux (for our particular geometry, see Fig. 2) 
28.8 \%, (15.4 \% + 13.4 \%) goes directly to observer  and 25.6 \% 
after scattering in the converging Compton cloud.
But a smaller flux of the detected  photons 
 emerges after  multiple scattering and  they form the hard tail of 
the resulting spectrum (Fig. 5). 
 We remind the reader (see for details ST80, T94, and TMK97) that
 the photons undergoing multiple scatterings produce the 
specific space distribution (in accordance to the first space 
eigenfunction, see TZ98, Fig. 3) and that fraction of  photons undergoing 
multiple scattering in the plasma cloud, $f_{ms}$
is related to the expansion coefficient of the space source distribution
over  the first space eigenfunction. In the simplest case,  the uniform 
source distribution $f_{ms}$ is approximately  0.8. 
But  in our case when the source 
 photon distribution is produced by an external 
illumination of the converging inflow (CI) region  by the soft disk photons 
and the optical depth of the converging region is of order of 1 
($\tau\sim \dot m[1.5^{-1/2}-3^{-1/2}]\approx 1$),  
this fraction is  slightly less.  Here for the optical depth 
estimate we use formula (2) with an assumption that $r_{out}=3r_s$ and
$r_{in}=1.5r_s$ being as a photon bending radius (TZ98).  

Thus the fraction of photons undergoing multiple scattering and detected 
by the observer with respect those photons  undergoing a few scattering 
and  detected   or comes directly to the observer from the disk is 
estimated as follows
\begin{equation}
f\ltorder{{0.26\cdot 0.8}\over{0.26\cdot 0.2 + 0.29}}=0.6.
\end{equation}
 
The exact number of these photons can  be determined 
through the fitting  of the simulated resulting spectrum to the 
 test spectrum, composed of two components:       
a Planckian spectrum with a fixed temperature of the soft photons and       
convolution of some fraction $f$ of      
this component, with the CI upscattering  Green's 
function. It is easy to prove (using the property of the Green's function
to conserve the number of the scattering photons) that  $f$ is
exactly  the fraction of photons (with respect to the thermal component) 
multiply scattered in the converging
inflow. As shown in \S 3,  $f=0.48$ (for our case)
and  the flux of the escaping hard photons is  
large enough to explain the observations.   
We remind the reader that the empirically determined fraction 
are 0.32 and 0.72  for GRO J1655-40 and 
GRS 1915+105 respectively (ShT98) and {\it thus  
Comptonization by the converging inflow is consistent with the observations}.

It is worth noting that the number of the photons forming the very hard 
tail (for energies $E\gg kT_c$)
\begin{equation}
N_{hard}(E)=\int_E^{\infty}I(E_0){{dE_0}\over{E_0}}
\end{equation}
is much less than the number of photons in the thermal component $N_{th}$, i.e.
\begin{equation}
{{N_{hard}(E)}\over{N_{th}}}< f(E/2kT_c)^{-\alpha}\ll 1.
\end{equation}

For example, this ratio is   about  $8\cdot 10^{-2}$ 
for $E=10$ keV in the cases
of GRO J1655-40 and GRS 1915+105.
 In other words only 8\% 
of the photons 
seen as coming from the disk are effectively scattered off electrons 
within a few Schwarzschild radii ($2-3~ r_s$) and  form the extended power law 
tail of the spectrum. Inequality (11) can be readily proved using  a formula 
for the hard tail energy flux $I(E)$ (Eq. 10) presented in the convolution form
(see  ShT98, Eq. 2).

In the end of this section we discuss a issue related with the illumination 
of the CI atmosphere by  the disk photons. 
For the beginning we assume that the disk emitting isotropically 
 is extended from the inner radius $r_{in}$ to the outer radius $r_{out}$ 
and the outer radius of the CI atmosphere is $r_{sp}$.
And further more we assume that $r_{sp}\leq r_{in}$. 

In fact, the disk flux intercepted by the
the CI atmosphere with radius $r_{sp}$ 
(see Fig. 2) is calculated through the integral  
\begin{equation}
F_{int}=\pi\int_{r_{in}}^{r_{out}}rdr\int_{-\pi/2}^{\pi/2}\cos{\varphi}
d\varphi \int^1_{\mu_{\ast}}(1-\mu^2)^{1/2}d\mu
\end{equation}
where $\mu_{\ast}=\sqrt{1-(r_{sp}/r)^2}$. 
In the case when the disk illuminates the slab  
(with the half-width $\Delta r=r_{in}=3r_s$ and the half-height $H$)
situated in the center~~$\mu_{\ast}=(r-r_{in})/\sqrt{(r-r_{in})^2+H^2}$.

The two internal integrals of Eq. (12) are calculated analytically.
The first one is equal to 2 and the second one is~~
$\theta_{\ast}-0.5\sin{2\theta_{\ast}}$, 
where $\theta_{\ast}=\arccos\mu_{\ast}$.  
 Even the triple integral of Eq. (12) 
can be  estimated analytically with accuracy better than 10\% 
for the spherical case with the aforementioned assumption 
that $r_{sp}\leq r_{in}$:    
\begin{equation}
F_{int}= {{2\pi}\over3} r_{sp}^3(r_{in}^{-1}-r_{out}^{-1}).
\end{equation}

In the case when the part of the disk is covered by the CI atmosphere 
the flux,  $F_{int}$ is calculated by integration from $r_{sp}$ to 
$r_{out}$ followed by adding $\pi^2(r_{sp}^2-r_{in}^2)$, 
the disk flux emitted inside the atmosphere.

The fraction of the disk photons illuminating the CI 
atmosphere is 
\begin{equation}
f_{ill}= {{F_{int}p_{sc}}\over{\pi^2(r^2_{out}-r^2_{in})}}.
\end{equation} 
Where $p_{sc}$ is  a probability of scattering of the disk photons 
in the CI atmosphere which is of order 
$1-e^{-\tau}$. 

The factor $f$,  used in the spectral fitting can be estimated from here
\begin{equation}
f\sim {{f_{ill} f_{ms}A}\over{(1-f_{ms})f_{ill}A+\mu(1-f_{ill})}}
\end{equation}
where $\mu$ is the cosine of inclination angle of the disk and 
$A$ is a spherical albedo of the CI atmosphere.

The quantity $A$  can be estimated as follows (e.g. Sobolev 1975)  
\begin{equation}
A=1-{1\over{3\tau/4+1}}.
\end{equation}
This formula is obtained in the Eddington approximation 
with an assumption of isotropic scattering at any event and 
the pure absorptive inner boundary. Despite of  the Eddington approximation
the albedo formula (16) provides an accuracy of order 10\% for any optical 
depth. For optical depth 1,  $A=0.42$ which  is slightly higher 
than  value obtained in our Monte Carlo simulations,
 $A_{MC}= 25.7/(25.7+45.6)=0.36$ (see above) because of the photon bending.

Now we present the numerical estimates of the factor $f$ for two
different cases. The first one related with our Monte Carlo simulations
when $p_{sc}\sim 1-e^{-1}=0.63$, $f_{ms}=0.8$, $A=0.42$ and we assume 
that $\mu\sim 1$. Then we get that $f\sim 0.5$ which is very close to 
that we get from the spectral fitting of the Monte Carlo simulated 
spectrum. It is worth noting here that in the limit of the high optical
depth (or mass accretion rate) $f\sim f_{ms}/(1-f_{ms})$ because   
$p_{sc}$, $f_{ill}$, and $A$ converge to one. 
Thus in this particular case the factor $f$ can be of order of one or more.

For another case we assume that the soft radiation
comes from the extended disk with the inner radius $r_{in}=3 r_s$ and
$r_{out}=11 r_s$ which is the size of the disk emission region 
related with the color temperature of 0.85 keV 
(for  details of this model see Borozdin et al. 1998)
and the inclination direction cosine $\mu=0.5$. Also we use the above values of
$p_{sc}\sim 0.63$, $f_{ms}\sim 0.8$, $A\sim 0.42$. 
Using Eqs (12), (14-15) we obtain that $f=0.68 \cdot 10^{-2}$
if we consider the illumination of the sphere situated in the center with
 $r_{sp}=3 r_s$. The factor $f$ is $1.7\cdot 10^{-2}$ 
in the case of the illumination of the  slab 
with the  half-height, $H=3r_s$. 

To conclude this section we emphasize once again
that the Comptonizing region must overlap with the 
inner region of the disk, in order to get the high fraction of 
the  seed photons from the disk (see ShT98).
But from the other hand the high fraction, $f$ 
(of order 10 \% and higher) can be also obtained
if the innermost  region of the disk is puffed up to heights of 
 order of 3 $r_s$ and the disk is almost seen at edge on. 
More details of this discussion and comparison with the RXTE observations
would be presented in our incoming paper (Borozdin et al. 1998).

\section{Conclusions}      
      
We  show that spectra in the soft and hard states are formed by       
upscattering of the disk soft photons by the matter (electrons)       
moving in the strong gravitational field.       
 
The presence of the event horizon as well as the behaviour of the      
null geodesics in its vicinity  largely determine the dependence      
of the spectral index  on the flow parameters.      
The efficiency of gravitational to radiative       
energy conversion is of order 5\% or less in the disk around black hole       
(e.g.  SS73; Narayan, Barret \& McClintock 1997).  
Thus  it is natural to expect      
that radiation pressure cannot hold back the matter which is  almost       
in free-fall. But the shape of the spectrum (the spectral index and       
the position of the high energy cutoff) depends on the temperature of the       
bulk motion atmosphere, which is determined by the efficiency       
of the cooling of the atmosphere by the flux coming from the disk      
(see \S 5 here and CT95;  Titarchuk, Lapidus \& Muslimov 1998).      
In the hard state the most gravitational energy is released out of the disk     
(see possible scenarios in CT95) and the bulk motion zone is getting very       
hot [because there is no enough cooling agents (soft disk photons) to cool      
down the environment]. The plasma temperature is of the order of 50 keV. 
 
In this case, the contribution of the outer part of the bulk motion atmosphere  
in the resulting spectrum is significant because of the large volume of this  
region  and of the high thermal velocities. 
In fact, the hard photons are produced mainly by effective scatterings of  
the disk soft photons on electrons at angles near $\pi$ 
(Hua \& Titarchuk 1995, Fig. 2).  
But in the outer layer the gravitational bending and redshift is less  
pronounced and thus the effect of General Relativity on the spectral index  
is smaller for larger values of electron temperature as shown in table  
2. Indeed, the spectrum is practically identical  to the standard  
thermal Comptonization spectrum (see Fig. 6).  
 
When the mass accretion rate in the disk increases,       
the plasma temperature goes down  to such a level (of order 10 keV or less)  
that only the free-falling electrons  transfer  their momentum to the      
soft-photons  producing the power-law component extending to energies       
comparable to the kinetic energy of electrons in the converging inflow,       
i.e. of order $m_ec^2$.         
 In this case, the spectrum observed at infinity      
consists of a soft component produced by those input  photons that      
escape after a few scattering without any significant energy change      
and of an hard component (described by  a power law at the energies  
much higher than the characteristic energy of the disk soft photons) 
 produced by the photons that undergo significant upscattering.      
 
The luminosity of the power-law component is relatively small compared      
to that of the soft component. The power law is quite steep due to  
the strong photon bending and the gravitational redshift  
near the black hole (within 1.5-2 Schwarzschild radii).    
We demonstrate through our Monte Carlo simulations 
 how many photons {\it escape} from the central region, 
when a given number of photons are injected in the Monte-Carlo simulations
and that fraction is perfectly consistent with the recent high energy 
observations (ShT98).   

We also show that our Monte-Carlo spectra are fitted quite well by the 
the analytical bulk motion Comptonization model (BMC) which allows 
one to use this model for the fitting of the real observational data.
The analytical model (BMC) contains only three parameters. 
 Two of them are the color temperature of the seed photons $T_c$ and 
the relative weight of the soft component in the entire spectrum, $(1-f)$
which are the main characteristics of the soft component.
Thus it is getting possible  to determine 
 the color temperature dependence on the energy flux in  the soft  
 energy band.
The disk origin of the soft seed photons can be confirmed or refuted
by this dependence (see  details in Borozdin et al. 1998).

We demonstrate using our Monte-Carlo simulation 
that   the main features of the simulated spectra  
( a spectral index of $1.5$ and an high energy       
cutoff around 500 keV) are consistent with the observed X-ray spectra        
in the high state of BHS (e.g. Grove et al. 1998).
The predicted high energy cutoff 
(assuming a relativistic  Maxwellian electron distribution) 
is in the energy range between 200 and 400 keV (see Figs 4-5).
It is slightly less than that observed in the superluminal source
GRO J1655 by OSSE  (Grove et al. 1998, table 2 and Fig. 2). 
It is worth noting  that the break energy is quite sensitive 
to the electron distribution. 
 In future work, we will show 
that the observed spectrum (including the position of 
the high energy break) can indeed be accounted 
for by the converging model quantitatively using other electron distributions.
In fact, we plan to investigate the possibility of non-thermal tails
to the electron distribution in accreting systems and the expected 
radiation signatures in the hard X-ray band.
   
{\it Thus our results of the Monte Carlo simulations 
strongly support the idea   that the bulk motion      
Comptonization might be responsible for  the extended       
power-law spectra seen in the black-hole X-ray sources in their soft state. 
And the hard-soft states transition is regulated by the plasma temperature 
of the converging inflow into a black hole.}

\acknowledgements{ L.T. would like to acknowledge support from NASA grant 
NAG5-4965, Chris Shrader, Wei Cui and Jean Swank for  encouragement and 
discussion and  Michael Revnivtsev and Konstantin Borozdin for help to fit   
the Monte Carlo spectra to XSPEC analytical models.  Also P.L. and L.T. 
 would like to acknowledge Paolo Goldoni for discussion and Jacques Paul 
for support of this work. Particularly, we thank anonymous referee for
interesting questions and helpful suggestions that significantly improved the
paper presentation.}

\clearpage 
 
\begin{figure} 
\caption{The left-hand panel is an artists 
conception of an accretion disk surrounding a black hole. 
The approximate sites from which the soft-seed radiation is produced, 
the more general soft radiation which is directly viewed by the observer
(thick solid line),  
and the hard-radiation emission ((thin solid line) are indicated. 
Also, the dotted lines indicate the 
scale and location of the bulk-motion scattering site. The schematic  
diagram in right-hand panel, depicts a typical sequence of scatterings  
within the bulk-motion inflow region. Some details are given in the text.} 
\label{Fig. 1} 
\end{figure}              
 
\begin{figure} 
%\centerline{\epsfig{figure=fig1f98.ps,height=50mm}} 
\caption{Scheme of the geometrical setup used 
in the Monte-Carlo simulations. We vary the accretion sphere radius 
$r_{out}$ from  3 $r_s$ to 20 $r_s$ in our simulations (see text).} 
\label{Fig. 2} 
\end{figure}              
 
\begin{figure} 
\caption{The Monte Carlo photon spectrum (see below Fig. 5) 
 is fitted to the test (analytical) spectrum, composed of two components:       
a Planckian spectrum with a fixed temperature of the soft photons and       
convolution of some fraction of      
this component with the CI upscattering  Green's function.} 
\label{Fig. 3} 
\end{figure}              
      
\begin{figure} 
\caption{Emergent Comptonized photon 
spectrum in the general relativistic case:       
in this simulation, the cloud has a temperature of 5 keV       
and $\dot m$ is equal to 2.       
The best power law fit also shown has a  a photon number index of 2.9
(or  the spectral index $\alpha=1.9$).} 
\label{Fig. 4} 
\end{figure}              
 
\begin{figure} 
\caption{Emergent Comptonized photon 
spectrum in the general relativistic case:       
the physical conditions of the cloud are now: $kT_e =$ 20 keV, 
$\dot m = 4$.     
The best power law fit also shown has a photon number index of 2.5 
(or a spectral index $\alpha= 1.5$).} 
\label{Fig. 5} 
\end{figure}              
 
\begin{figure} 
\caption{Emergent Comptonized photon 
spectrum in the general relativistic case       
with this time $kT_e =$ 50 keV, and $\dot m = 2$.       
The high energy part of the spectrum is compared       
to the thermal Comptonization spectrum [Hua \& Titarchuk (1995), Eq. 6].} 
\label{Fig. 6} 
\end{figure}              
 
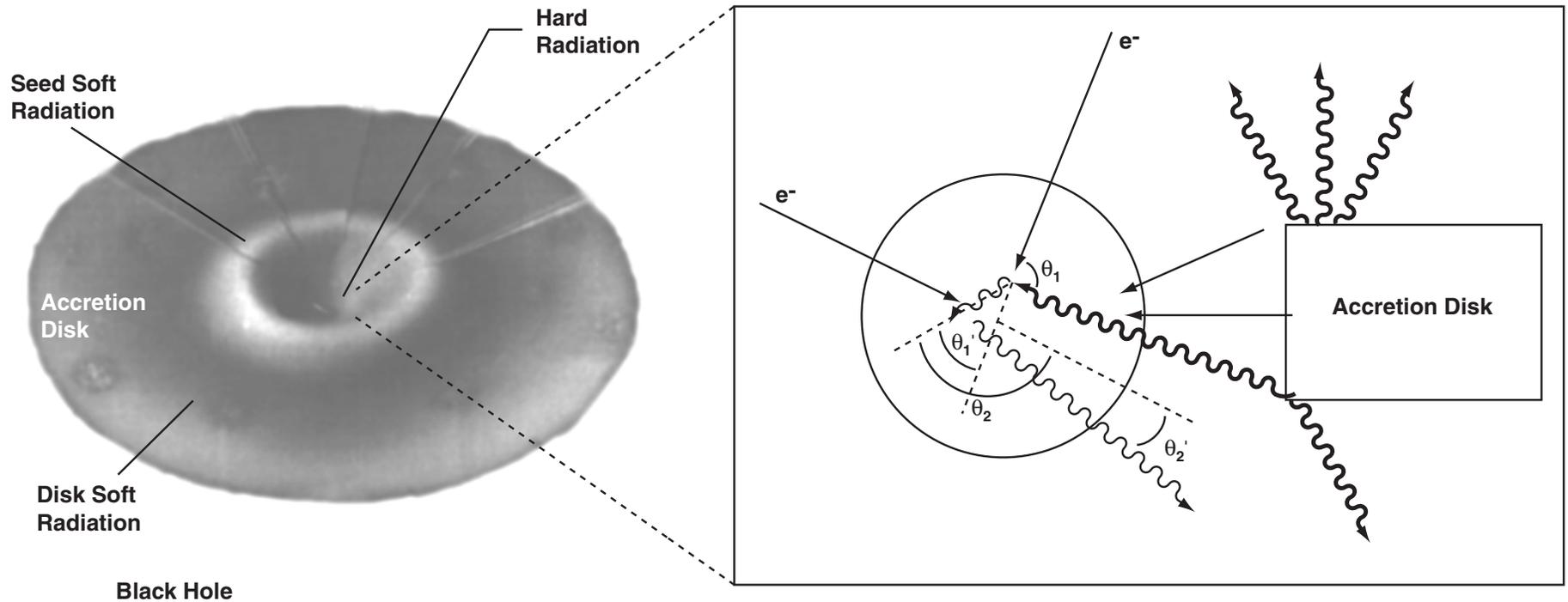
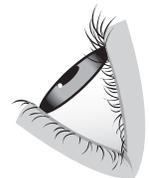
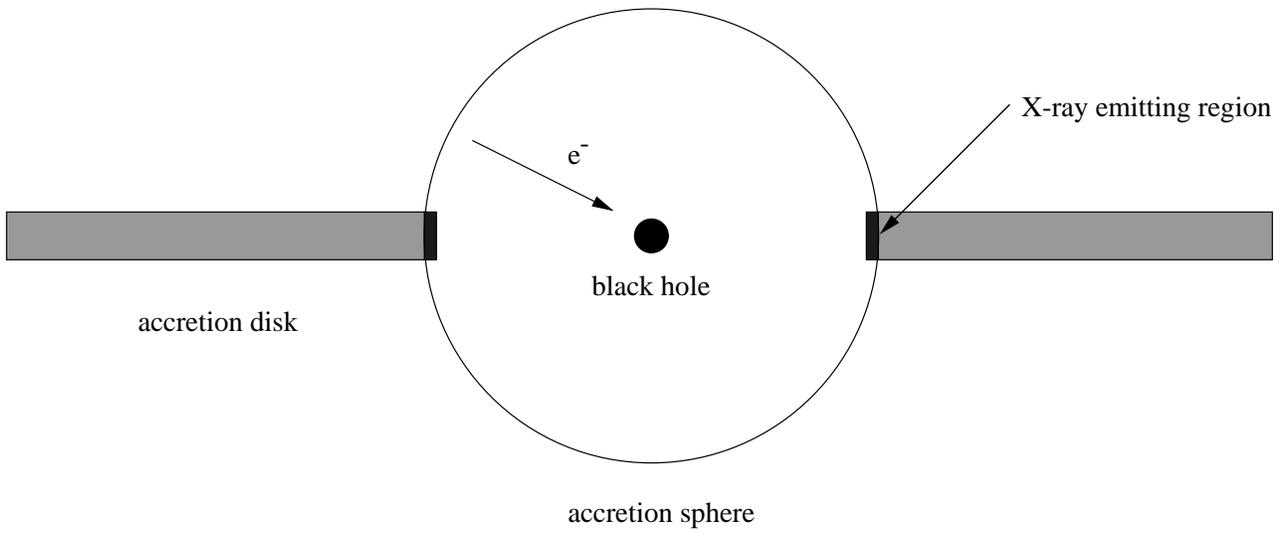
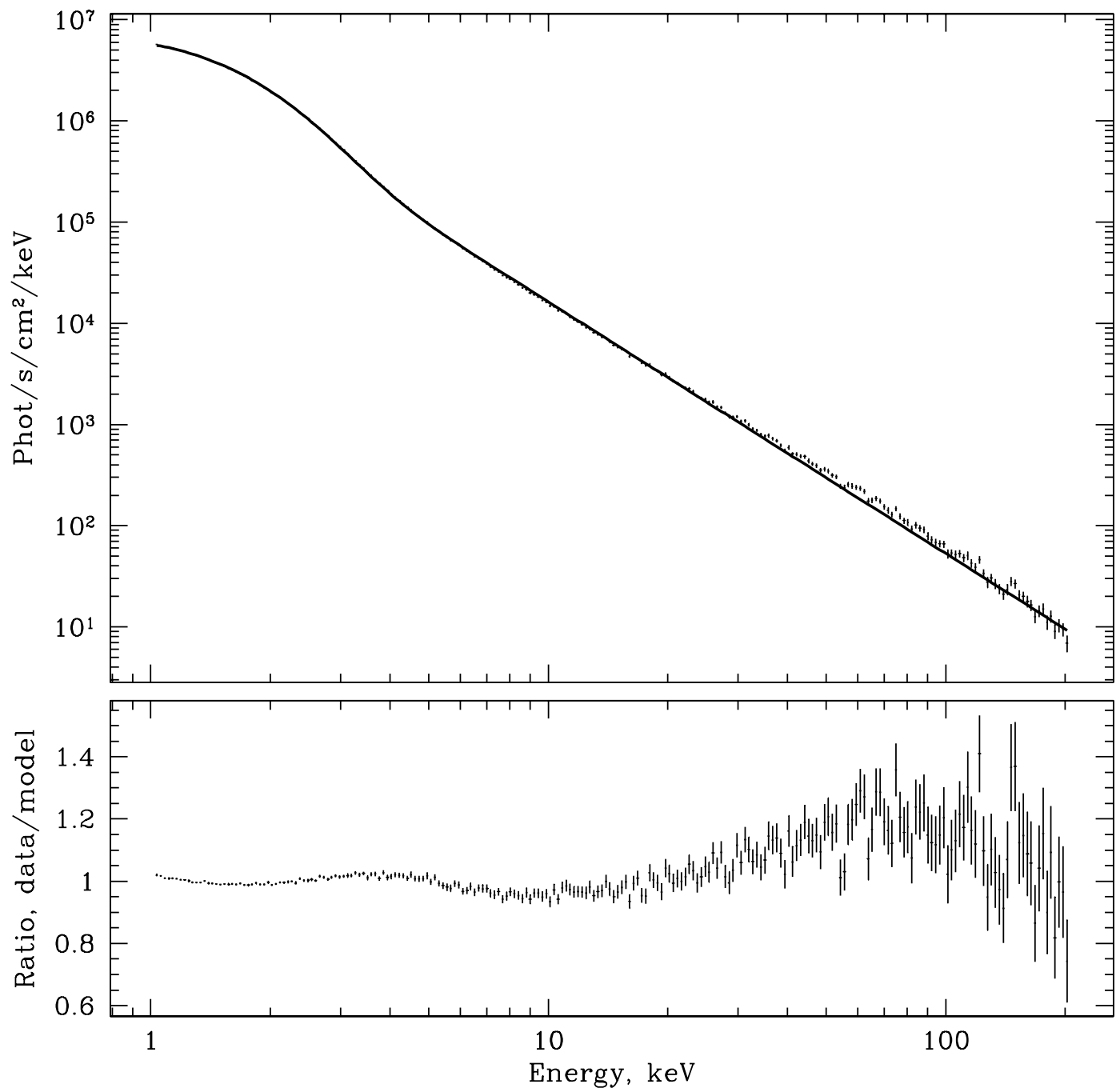
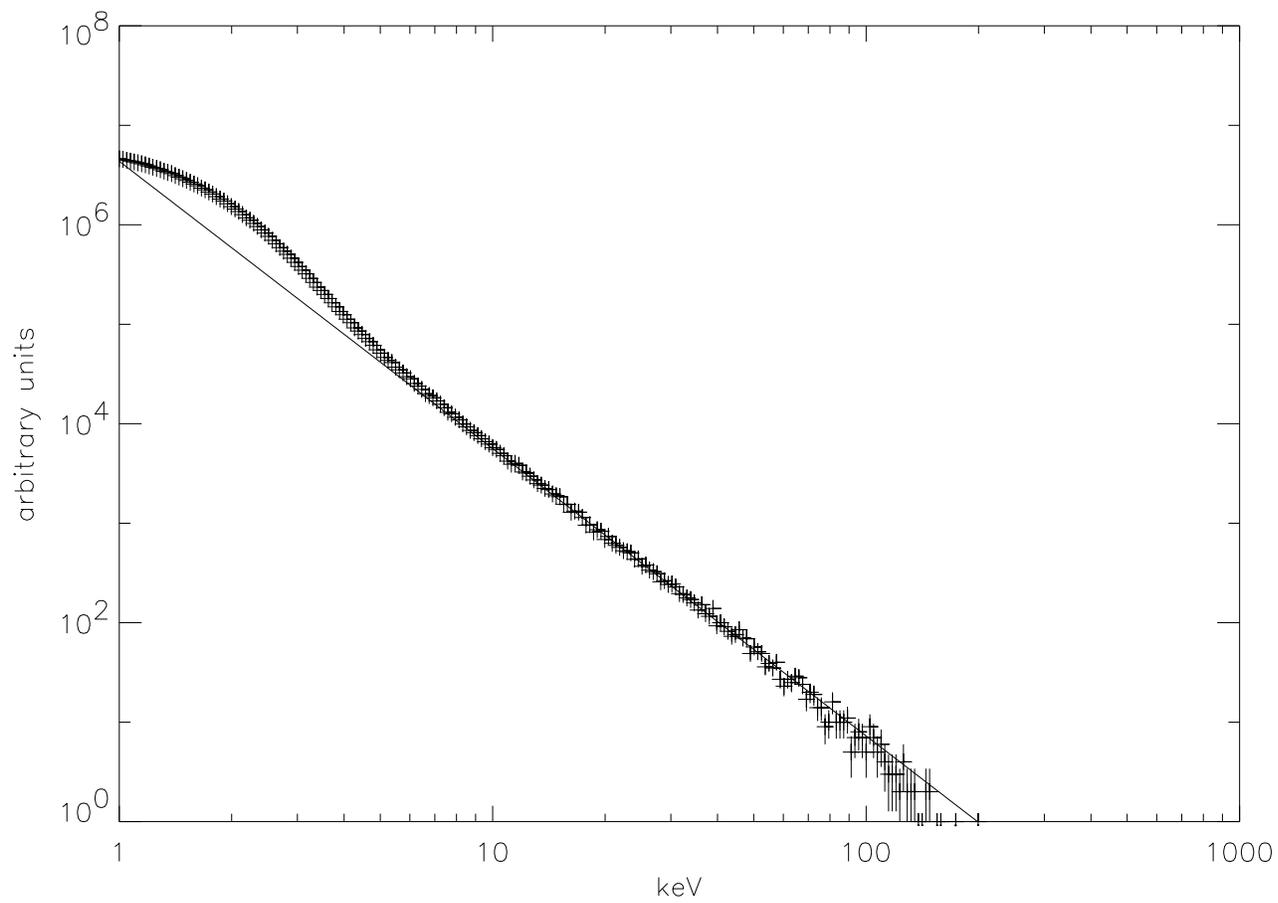
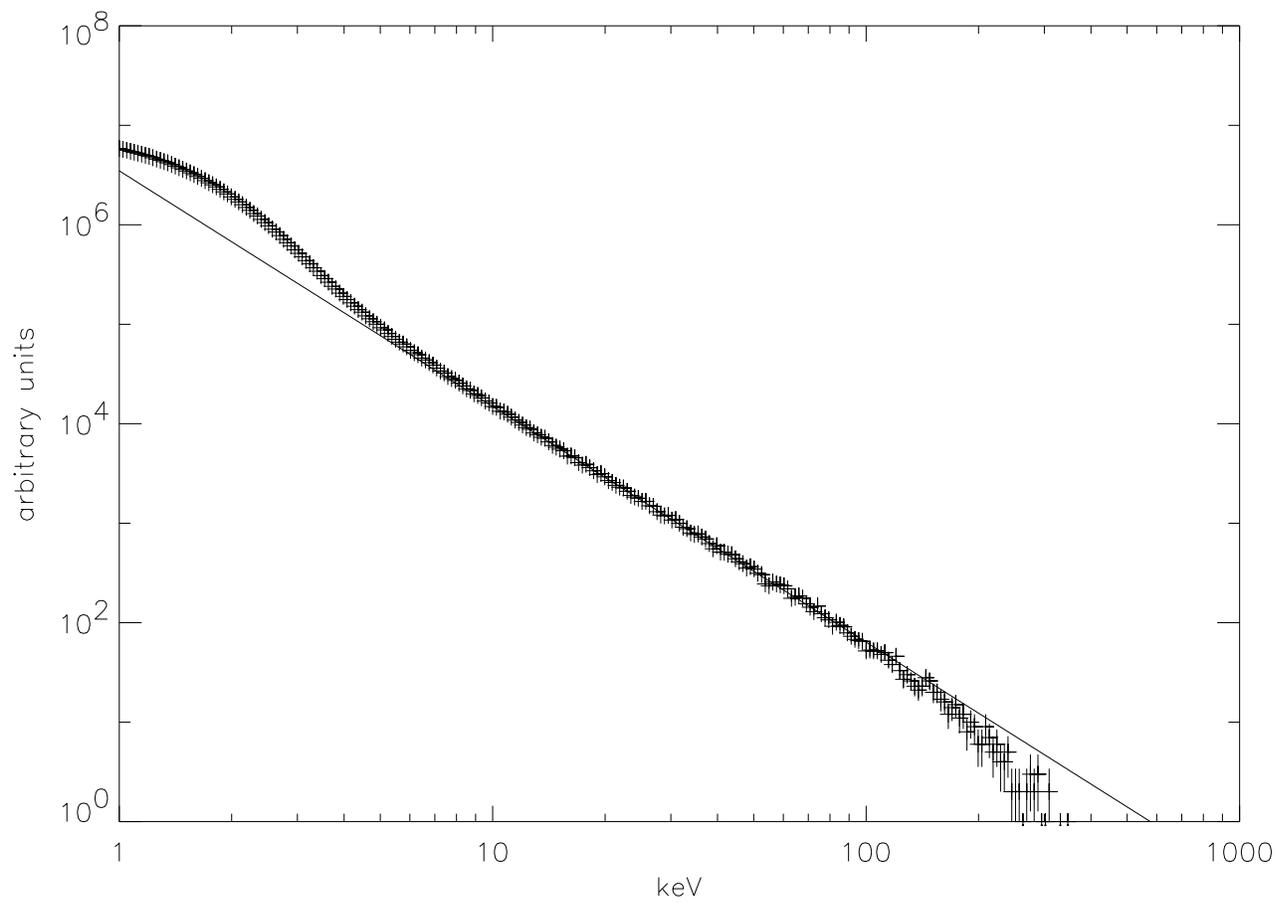
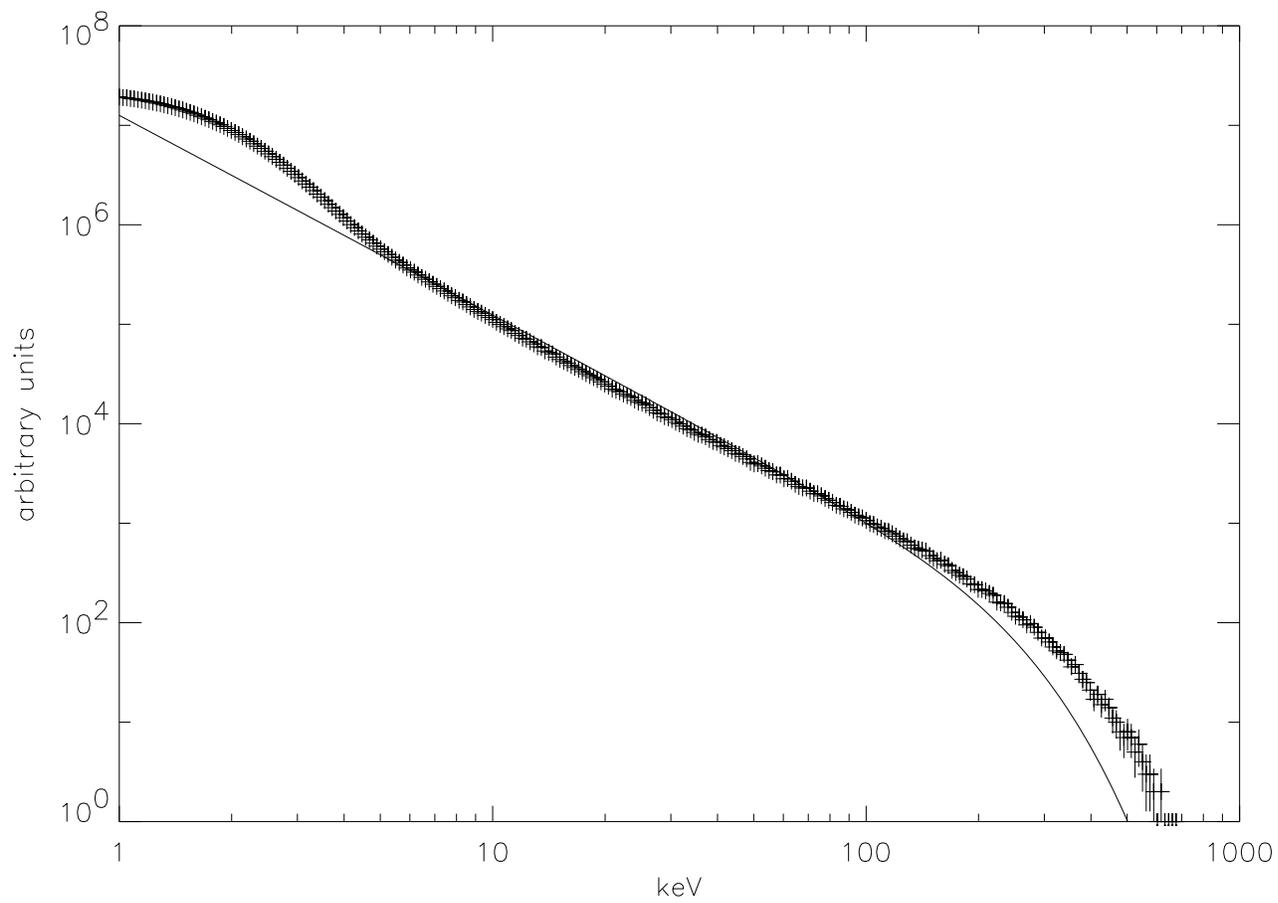
\end{document}